\documentclass[12pt]{article}

\usepackage{amsmath}
\usepackage{a4wide}
\usepackage{amsmath,amssymb}

\begin{document}

\title {\large Supplement to: Code Spectrum and Reliability
Function: Binary Symmetric Channel}

\author {Marat V. Burnashev \\
Institute for Information Transmission Problems, \\
Russian Academy of Sciences,
Moscow, Russia \\
Email: burn@iitp.ru}

\date{}

\maketitle

A much simpler proof of Theorem 1 from \cite{Bur6} is presented
below, using notation and formulas numeration of \cite{Bur6}. The
text below replaces the subsection {\bf General case} from \S 4
of \cite[p. 11]{Bur6}.

{\bf General case}. In the general case for some $\omega$ we are
interested in a pairs
$(\mbox{\boldmath $x$}_{i},\mbox{\boldmath $x$}_{j})$ with
$d_{ij} = \omega n$. But there may exist a pairs
$(\mbox{\boldmath $x$}_{k},\mbox{\boldmath $x$}_{l})$ with
$d_{kl} < \omega n$. Using the ``cleaning'' procedure \cite{Bur5}
we show that the influence of such pairs
$(\mbox{\boldmath $x$}_{k},\mbox{\boldmath $x$}_{l})$ on the value
$P_{\rm e}$ is not large. It will allow us to reduce the general
case to the model one.

Note that if
\begin{equation}
\frac{1}{n}\log X_{\rm max}(t,\omega) = o(1)\,, \quad
n \to \infty\,, \tag{S.1}
\end{equation}
then from (27) and (28) we get
\begin{equation}
\begin{gathered}
\frac{1}{n}\log \frac{1}{P_{\rm e}} \leq -\log q + \\
+ \min_{0 \leq t \leq 1}\min_{\omega}\left\{t\log \frac{q}{p} -
\omega - (1-\omega)h_{2}\left(\frac{1}{2} -\frac{1-2t}
{2(1-\omega)}\right) - b(\omega)\right\} + o(1)\,, \tag{S.2}
\end{gathered}
\end{equation}
where $b(\omega) = n^{-1}\log B_{\omega n}$.

The minimum over $t$ in the right-hand side of (S.2) is attained
when
\begin{equation}
\begin{gathered}
t(\omega) = \frac{\omega}{2}+(1-\omega)p\,, \tag{S.3}
\end{gathered}
\end{equation}
and then (S.2) takes the form
\begin{equation}
\begin{gathered}
\frac{1}{n}\log \frac{1}{P_{\rm e}} \leq \min_{\omega}f(\omega) +
o(1)\,, \qquad f(\omega) = \frac{\omega}{2}\log \frac{1}{4pq}-
b(\omega)\,. \tag{S.4}
\end{gathered}
\end{equation}
Let $f(\omega)$ attains its minimum (over all $\omega$) at some
$\omega_{0}$. By definition we have for any $\omega$
\begin{equation}
\frac{\omega_{0}}{2}\log \frac{1}{4pq} - b(\omega_{0}) \leq
\frac{\omega}{2}\log \frac{1}{4pq} -b(\omega)\,. \tag{S.5}
\end{equation}

To avoid a superfluous awkwardness, we omit the remaining term
$o(1)$ in the Theorem 2. Then there exists $\omega$ such that
$\omega \leq G(\alpha,\tau)$ and
$b(\omega) \geq \mu(R,\alpha,\omega)$.
Denote $\omega^{*}$ the smallest $\omega \leq G(\alpha,\tau)$ for
which we have $b(\omega) \geq \mu(R,\alpha,\omega)$.

We call $(\mbox{\boldmath $x$}_{i},\mbox{\boldmath $x$}_{j})$ a
$\omega$--pair if $d(\mbox{\boldmath $x$}_{i},
\mbox{\boldmath $x$}_{j}) = \omega n$. Then the total number of
$\omega$--pairs equals $M2^{nb(\omega)}$. We use $t= t(\omega_{0})$
from (S.3), and say that a point $\mbox{\boldmath $y$}$ is
$\omega$--covered if there exists a $\omega$--pair
$(\mbox{\boldmath $x$}_{i},\mbox{\boldmath $x$}_{j})$
such that $d(\mbox{\boldmath $x$}_{i},\mbox{\boldmath $y$}) =
d(\mbox{\boldmath $x$}_{j},\mbox{\boldmath $y$}) = tn$. Then there
are $M2^{nb(\omega)}Z(t(\omega_{0}),\omega)$ $\omega$--covered
points $\mbox{\boldmath $y$}$ (taking into account the covering
multiplicities). Introduce the set ${\mathbf Y}(\omega)$ of all
$\omega$--covered points $\mbox{\boldmath $y$}$.
We set a small $\varepsilon > 0$ and perform a cleaning procedure.
Consider the set ${\mathbf Y}(\omega_{0})$ and exclude from it all
points $\mbox{\boldmath $y$}$ that are also $\omega$--covered for
any $\omega$ such that $|\omega - \omega_{0}| \geq \varepsilon$,
i.e. consider the set of all $\omega_{0}$--covered points
$\mbox{\boldmath $y$}$ which are not $\omega$--covered for any
$\omega$ such that $|\omega - \omega_{0}| \geq \varepsilon$:
\begin{equation}
\begin{gathered}
{\mathbf Y}'(\omega_{0}) = {\mathbf Y}(\omega_{0}) \setminus
\bigcup_{|\omega - \omega_{0}| \geq \varepsilon}
{\mathbf Y}(\omega). \tag{S.6}
\end{gathered}
\end{equation}

Each point $\mbox{\boldmath $y$} \in {\mathbf Y}'(\omega_{0})$ can
be $\omega$--covered only if
$|\omega - \omega_{0}| < \varepsilon$. We show that for an
appropriate $\varepsilon$ both sets
${\mathbf Y}(\omega_{0})$ and ${\mathbf Y}'(\omega_{0})$ have
essentially the same cardinalities. Each $\omega$--pair
$(\mbox{\boldmath $x$}_{i},\mbox{\boldmath $x$}_{j})$
$\omega$--covers the set ${\mathbf Z}_{ij}(t,\omega)$ with the
cardinality $Z(t,\omega)$. We compare
the values $\sum\limits_{|\omega - \omega_{0}| \geq \varepsilon}
2^{nb(\omega)}Z(t,\omega)$ and $2^{nb(\omega_{0})}Z(t,\omega_{0})$
(see (S.6)). For that purpose consider the function
\begin{equation}
f(\omega) = \frac{1}{n}\log \frac{2^{nb(\omega)}Z(t,\omega)}
{2^{nb(\omega_{0})}Z(t,\omega_{0})} = b(\omega) - b(\omega_{0}) +
u(t,\omega) - u(t,\omega_{0})\,, \tag{S.7}
\end{equation}
where
$$
u(t,\omega) = \omega + (1-\omega)h_{2}\left[\frac{1}{2} -
\frac{(1-\omega_{0})(1-2p)}{2(1-\omega)}\right]\,.
$$
Due to (S.5) we have
$b(\omega) - b(\omega_{0}) \leq (\omega_{0} - \omega)[\log(4pq)]/2$,
and then for the function $f(\omega)$ from (S.7) we get
\begin{equation}
\begin{gathered}
f(\omega) \leq v(\omega) = -\left(\frac{1}{2}-p\right)
(\omega_{0} - \omega)\log\frac{q}{p} + \\
+(1-\omega)\left[h_{2}\left(\frac{1}{2}-
\frac{(1-2p)(1-\omega_{0})}{2(1-\omega)}\right)-h_{2}(p)\right]\,, \\
v' = \frac{1}{2}\log \frac{1}{4pq} + \frac{1}{2}\log
\left[1-\frac{(1-2p)^{2}
(1-\omega_{0})^{2}}{(1-\omega)^{2}}\right]\,, \tag{S.9}\\
v'' = - \frac{(1-2p)^{2}(1-\omega_{0})^{2}\log_{2}e}
{(1-\lambda)[(1-\omega)^{2}-(1-2p)^{2}(1-\omega_{0})^{2}]} <
- \frac{(1-2p)^{2}}{3}\,.
\end{gathered}
\end{equation}
Since $v(\omega_{0}) = v'(\omega_{0}) = 0$, then for any $\omega$
we have
$$
f(\omega) \leq v(\omega) < - \frac{(1-2p)^{2}}{6}
(\omega_{0} - \omega)^{2}\,.
$$
Now after simple calculations we have
$$
\begin{gathered}
\sum\limits_{|\omega - \omega_{0}| \geq \varepsilon}
2^{nb(\omega)}Z(t,\omega)\Big/\left[
2^{nb(\omega_{0})}Z(t,\omega_{0})\right] =
\sum\limits_{|\omega - \omega_{0}| \geq \varepsilon}
2^{nf(\omega)} \leq \\
\leq 2\sum\limits_{\omega - \omega_{0} \geq \varepsilon}2^{
- (1-2p)^{2}(\omega_{0} - \omega)^{2}n/6}
= 2\sum\limits_{i \geq \varepsilon n} 2^{-(1-2p)^{2}i^{2}/(6n)}
\leq \\
\leq 2\left[1+\frac{3}{(1-2p)^{2}\varepsilon}\right]
e^{-(1-2p)^{2}\varepsilon^{2}n/6} \leq \frac{6n^{-1/6}}{1-2p}\,,
\end{gathered}
$$
if we set
$$
\varepsilon = \frac{2\sqrt{\ln n}}{(1-2p)\sqrt{n}}\,.
$$
Therefore for $n^{1/6} \geq 12/(1-2p)$ we get
$$
2^{nb(\omega_{0})}Z(t,\omega_{0}) -
\sum\limits_{|\omega - \omega_{0}| \geq \varepsilon}
2^{nb(\omega)}Z(t,\omega) \geq
\frac{1}{2}2^{nb(\omega_{0})}Z(t,\omega_{0})\,.
$$
In other words, all points
$\mbox{\boldmath $y$} \in {\mathbf Y}'(\omega_{0})$ are, in total,
$\omega$--covered, at least, $2^{nb(\omega_{0})}Z(t,\omega_{0})/2$
times, and, moreover, each point
$\mbox{\boldmath $y$} \in {\mathbf Y}'(\omega_{0})$ can be
$\omega$--covered only if $|\omega - \omega_{0}| < \varepsilon$.
Due to the formula (S.9) it means that the cardinalities of the sets
${\mathbf Y}(\omega_{0})$ and ${\mathbf Y}'(\omega_{0})$ have
equal exponential order.

For each point $\mbox{\boldmath $y$} \in {\mathbf Y}'(\omega_{0})$
consider the set ${\mathbf X}_{t}(\mbox{\boldmath $y$})$ defined in
(19), i.e. the set of all codewords
$\{\mbox{\boldmath $x$}_{i}\}$ such that
$d(\mbox{\boldmath $x$}_{i},\mbox{\boldmath $y$}) = t(\omega_{0})n$.
The codewords from ${\mathbf X}_{t}(\mbox{\boldmath $y$})$ satisfy
also the condition $|d(\mbox{\boldmath $x$}_{i},
\mbox{\boldmath $x$}_{j})- \omega_{0}n| \leq \varepsilon n$,
i.e. the set ${\mathbf X}_{t}(\mbox{\boldmath $y$})$ constitutes
almost a simplex. It is clear that the number
$\left|{\mathbf X}_{t}(\mbox{\boldmath $y$})\right|$ of such
codewords is not exponential on $n$, i.e.
\begin{equation}
\log \left|{\mathbf X}_{t}(\mbox{\boldmath $y$})\right| = o(n)\,,
\quad \mbox{\boldmath $y$} \in {\mathbf Y}'(\omega_{0})\,,
\qquad n \to \infty\,.  \tag{S.9}
\end{equation}
For accurateness the formula (S.9) is proved below. It follows from
(S.9) that the condition (S.1) is satisfied with
$X_{\rm max}(t,\omega_{0}) =
\max\limits_{i,\mbox{\boldmath $y$} \in {\mathbf Y}'(\omega_{0})}
\left|{\mathbf X}_{i}(\mbox{\boldmath $y$},t,\omega_{0})\right|$
(cf. (25)). Using the upper bound (S.4) and the inequality (S.5)
we get
\begin{equation}
\begin{gathered}
\frac{1}{n}\log \frac{1}{P_{\rm e}} \leq f(\omega_{0}) + o(1) \leq
f(\omega^{*}) + o(1) \leq \max_{\omega \leq
G(\alpha,\tau)}g(\omega) + o(1)\,, \tag{S.10}  \\
g(\omega) = \frac{\omega}{2}\log \frac{1}{4pq} -
\mu(R,\alpha,\omega)\,,
\end{gathered}
\end{equation}
from which the desired upper bound (11) follows.

It remains us to prove the relation (S.9). If
$\omega^{*} \geq \omega_{1}$ then (S.9) immediately follows from
\cite[proposition 4]{Bur6}. In the general case (S.9) follows from
the lemma.

{L e m m a}. {\it Let ${\cal C} = \{\mbox{\boldmath $x$}_{1},
\ldots,\mbox{\boldmath $x$}_{M}\}$ be a code
such that for some $\omega$ the relation holds
$$
\max\limits_{i \neq j}\left|d(\mbox{\boldmath $x$}_{i},
\mbox{\boldmath $x$}_{j}) - \omega n\right| = o(n),\qquad
n \to \infty\,.
$$
Then}
\begin{equation}
\begin{gathered}
n^{-1}\ln M \to 0\,, \qquad n \to \infty\,. \tag{S.11}
\end{gathered}
\end{equation}

{P r o o f}. If $\mbox{\boldmath $x$}_{i},\mbox{\boldmath $x$}_{j}$
are binary codewords then for their Hamming and Euclidean distances
we have $d_{\rm H}(\mbox{\boldmath $x$}_{i},
\mbox{\boldmath $x$}_{j}) = \|\mbox{\boldmath $x$}_{i} -
\mbox{\boldmath $x$}_{j}\|^{2}$. Without loss of generality we may
assume that all codewords $\{\mbox{\boldmath $x$}_{i}\}$ have the
same Hamming weight $An$. Then a binary code
$\{\mbox{\boldmath $x$}_{i}\}$ of the length $n$ can be
considered as an Euclidean code $\{\mbox{\boldmath $x$}_{i}\}
\subset S^n(\sqrt{An})$. For the Euclidean case the relation (S.11)
has been proved in \cite[Lemma 2]{Bur7}. $\qquad \blacktriangle$

It finishes the upper bound (11) proof. $\qquad \blacktriangle$


\medskip

\begin{center} {\large \bf REFERENCES} \end{center}
\begin{enumerate}

\bibitem{Bur6}
{\it Burnashev M. V.} Code spectrum and reliability function: binary
symmetric channel // Probl. Inform. Transm. 2006. V. 42. ü 4.
P. 3--22; \\ also http://arxiv.org/cs.IT/0612032.
\bibitem{Bur5}
{\it Burnashev M. V.} Upper bound sharpening on reliability function
of binary \\ symmetric channel // Probl. Inform. Transm. 2005.
V. 41. No. 4. P. 3--22.
\bibitem{Bur7}
{\it Burnashev M. V.} Code spectrum and reliability function:
Gaussian channel // Probl. Inform. Transm. (in print).

\end{enumerate}
\end{document}